\newcommand{\PaperTitle}{A Measurement of Genuine Tor Traces for\\Realistic Website Fingerprinting}

\documentclass[10pt,sigconf,letterpaper,nonacm,natbib=false]{acmart}

\usepackage[false]{anonymous-acm}

\setcopyright{acmlicensed}
\copyrightyear{2024}
\acmYear{2024}
\acmDOI{XXXXXXX.XXXXXXX}

\usepackage[T1]{fontenc}
\usepackage[utf8]{inputenc}
\usepackage[american]{babel}
\usepackage{csquotes, xpatch} \usepackage{microtype}
\microtypecontext{spacing=nonfrench}

\RequirePackage[ backend=biber,
  sorting=nty,
  mincrossrefs=1000,
  maxnames=99,
  date=year,
  doi=true,
  eprint=true,
  isbn=false,
  style=acmnumeric, citestyle=numeric-comp,
  datamodel=acmdatamodel, ]{biblatex}
  \DeclareFieldFormat{eprint}{\textsc{eprint:} \url{#1}}
\AtEveryBibitem{
   \clearfield{pages}
   \clearname{editor}
   \clearlist{location}
   \clearfield{venue}
   \clearfield{eventyear}
   \clearlist{publisher}
}
\AtBeginBibliography{\footnotesize}
\usepackage{amsmath, amsthm, amsfonts, bm} \usepackage{balance} \usepackage{booktabs} \usepackage{multicol, multirow, tabularx} \usepackage{tabulary}
\usepackage{pdflscape}

\usepackage{enumitem} \setlist[itemize]{label=--,leftmargin=*,itemsep=1pt,topsep=0pt,parsep=0pt}
\setlist[enumerate]{leftmargin=*,itemsep=1pt,topsep=0pt,parsep=0pt}
\setlist[description]{font={\normalfont\itshape}}

\usepackage[flushleft]{threeparttable} \usepackage{caption} \usepackage{subcaption} \usepackage{doi} \usepackage{float} \usepackage{fontawesome}
\usepackage{graphicx} \usepackage{hyperref} 

\usepackage{grffile} \graphicspath{{./}}

\usepackage{placeins}

\usepackage[]{siunitx}
\usepackage{xcolor, colortbl} \usepackage{xfrac} \usepackage{xspace} \usepackage{xurl} 

\usepackage{listings}

\DeclareMathOperator{\sha}{sha256}
\DeclareMathOperator{\b64}{base64encode}
\newcommand\salt{\ensuremath \mathit{salt}}

\usepackage{tcolorbox}

\newcommand{\dataset}{GTT23}

\hypersetup{
     colorlinks=true,
     linkcolor=black,
     citecolor=black,
     filecolor=black,
     urlcolor=black,
     pdfauthor = {}, pdftitle = {\PaperTitle},
     pdfkeywords = {} }

\definecolor{background}{HTML}{EEEEEE}
\colorlet{punct}{blue}
\begin{document}

\renewcommand{\tableautorefname}{Table}
\renewcommand{\figureautorefname}{Fig.}
\renewcommand{\equationautorefname}{Eqn.}
\renewcommand{\sectionautorefname}{\S\kern-1.5pt}
\renewcommand{\subsectionautorefname}{\S\kern-1.5pt}
\renewcommand{\subsubsectionautorefname}{\S\kern-1.5pt}

\date{}

\title{\PaperTitle}

\author{{\rm Rob Jansen$^{\dagger}$ \hspace{3cm} Ryan Wails$^{\dagger\ddagger}$ \hspace{3cm} Aaron Johnson$^{\dagger}$}}
\affiliation{\institution{$^{\dagger}$U.S. Naval Research Laboratory, $^{\ddagger}$Georgetown University}
  \city{}
  \state{}
  \country{}
}

\begin{abstract}
    Website fingerprinting (WF) is a dangerous attack on web privacy because it
    enables an adversary to predict the website a user is visiting, despite the
    use of encryption, VPNs, or anonymizing networks such as Tor.
Previous WF work almost exclusively uses \emph{synthetic} datasets to
    evaluate the performance and estimate the feasibility of WF attacks despite
    evidence that synthetic data misrepresents the real world.
In this paper we present \dataset{}, the first dataset of \emph{genuine}
    Tor traces, which we obtain through a large-scale measurement of the Tor
    network and which is intended especially for WF.
    It represents real Tor user behavior better than any
    existing WF dataset, is larger than any existing WF dataset by at least an
    order of magnitude, and will help ground the future study of
    realistic WF attacks and defenses.
In a detailed evaluation, we survey 28 WF datasets published since 2008
    and compare their characteristics to those of \dataset{}. We
    discover common deficiencies of synthetic datasets that make them inferior
    to \dataset{} for drawing meaningful conclusions about the effectiveness of
    WF attacks directed at real Tor users.
We have made \dataset{} available to promote reproducible research and to
    help inspire new directions for future work.
\end{abstract}
 
\maketitle

\thispagestyle{plain}
\pagestyle{plain}

\section{Introduction}
\label{sec:intro}

\looseness-1
Website fingerprinting (WF) is a dangerous attack on web privacy because it
enables an adversary that can observe a user's outgoing connections to predict
the website the user is visiting~\autocite{SP:SSWRPQ02, PETS:Hintz02,
PETS:BLJL05, CCS:LibLev06, Herrmann2009}, even if those connections are
protected with encryption, virtual private networks (VPNs), or anonymizing
networks such as Tor~\autocite{USENIX:DinMatSyv04}. WF attacks are particularly
serious against Tor because they can break Tor's
anonymity~\autocite{Panchenko2011, CCS:CZJJ12, USENIX:WCNJG14, USENIX:HayDan16,
PoPETS:WanGol16, NDSS:PLPEZH16, CCS:SIJW18, NDSS:RPJvJ18, PoPETS:BLKD19,
PoPETS:OhSunHop19, CCS:SMRW19, PoPETS:RSMGW20, SP:Wang20, PoPETS:PulDah20,
PoPETS:OMRWH21, USENIX:CheJanTro22, PoPETS:JanWai23, USENIX:SJGLZX23,
SP:DYLZLXXW23, deng2025countmamba}.

Consistent with previous WF work, we consider an adversary that uses machine
learning (ML) to carry out WF attacks against Tor users from a vantage point on
the entry side of the Tor network. In using ML for WF, \emph{accurately labeled
traces} (i.e., examples) are required (1)\;to train WF classifiers to form
correct associations between encrypted traffic patterns and destination
websites, and (2)\;to evaluate classifier efficacy. A major problem for the
adversary is that information about the destination website needed to label the
traces is encrypted by Tor's onion routing scheme~\autocite{SP:SyvGolRee97} when
observed from the entry position. Thus, the adversary needs some other method to
collect labeled traces.

Through a survey of 28~WF datasets published since 2008 (see
\autoref{sec:evaluation}), we find that all but a \emph{single} prior study
consider an adversary that collects labeled traces using an automated browser
that programmatically fetches a set of selected webpages through
Tor~\autocite{tor-browser-selenium}. Such \textit{synthetic} datasets have been
criticized as unrepresentative of genuine Tor traffic along numerous
axes~\autocite{Perry2013, CCS:JAADG14, NDSS:RPJvJ18, PoPETS:JanWai23}, and their
use has led WF research to fall victim to several common ML evaluation
pitfalls such as the base rate
fallacy~\autocite{CCS:JAADG14, USENIX:CheJanTro22, USENIX:AQPWPWCR22}.

In an effort to address the serious limitations of synthetic WF datasets, a
recent study by Cherubin et al.~\autocite{USENIX:CheJanTro22} considers a WF
strategy in which the adversary uses a Tor exit relay to collect
\textit{genuine} traces, which can be observed and labeled by a relay in the
exit position. Genuine traces exhibit the real-world diversity in all factors
that might influence classifier performance, and they enable researchers to more
accurately evaluate the WF performance that we expect an adversary might
realistically attain. Unfortunately, this prior study was done in a
\emph{fully online} setting in order to avoid persistently storing genuine data
or trained WF classifiers. As a result, it is impossible to replicate their
results, and it is difficult to build on the methodology. Indeed, many later
works have continued to study WF using synthetically generated
datasets~\autocite{CCS:BaBoHo23, PoPETS:JanWai23, USENIX:SJGLZX23, SP:MHORHW23,
WPES:LBASM23, SP:DYLZLXXW23}.

In this paper we present \dataset{}, the first dataset of labeled
\emph{genuine Tor traces}. We describe a large-scale Tor relay measurement plan
that we designed to prioritize safety and privacy, which we developed through
consultation with our organization's Institutional Review Board and with the Tor
Research Safety Board~\autocite{trsb}. We executed our reviewed measurement
process to safely measure \num{13900621} circuits to \num{1142115} unique
destination domains and \num{68} unique destination server ports during a
13-week measurement period. We analyze \dataset{} and find that 96\% of the
measured circuits use ports 80, 8080, or 443 to first connect to a destination,
that most of the measured circuits carry fewer than 25 cells ($<$10.5 KB), and
that just a single circuit was measured for over 80\% of the measured domains.
Our analysis of \dataset{} helps demonstrate the high degree of traffic
diversity with which a WF adversary must contend when launching WF attacks in
the real world.

We further evaluate \dataset{} to compare its genuine characteristics to those
of existing synthetic WF datasets. First, we survey 28 WF datasets published
since 2008 and identify several common deficiencies of synthetic
datasets. We find that synthetic datasets are composed of a single
traffic type (web) using simplistic user models and static software tools while
focusing on website \emph{index pages} at uninformed base rates. Second, we conduct a
detailed analysis of the statistical disparities between \dataset{} and two
recent synthetic datasets that are specifically designed for more complex
\emph{website} fingerprinting wherein a website contains multiple accessible
webpages. We find that the circuit-length variation and website base rates are
still not reflected well in the synthetic datasets despite the improved modeling.

We conclude that, because \dataset{} contains genuine traces of websites
accessed by real Tor users at natural base rates, it is more realistic than any
existing synthetic dataset, and thus it will enable WF evaluations that more
accurately estimate real-world WF performance. We also note that, while \dataset{}
was designed to facilitate WF research, it may be useful for other research on Tor
traffic analysis, such as correlation attacks~\cite{nasr2018deepcorr} or malware
detection~\cite{dodia2022exposing}.

This dataset has been available to researchers upon request since
2024~\cite{gtt23_10620520}. 
This report contains details and analyses of the data to further the
understanding and use of \dataset{} and to promote the development of similar
datasets. It is a full version of the conference paper~\cite{gtt23-pam2026},
adding \autoref{circmeta}, \autoref{tab:http_size}, \autoref{app:ethics}, and some additional
details throughout. \section{Methodology}
\label{sec:methodology}

\subsection{Background}
\label{sec:methodology:background}

Tor~\cite{USENIX:DinMatSyv04} uses onion routing to anonymize TCP
connections on the Internet. The Tor network consists
of a globally distributed set of \emph{relays}. Each connection through
Tor to an outside server is sent through a three-hop \emph{circuit}. This
design is intended to prevent an adversary observing any single relay, or
one observing either the client or destination but not both, from being
able to identify both the source and destination of a connection.
In the Tor network today, there are currently over 8,500 relays and over
3 million daily users~\cite{tormetrics}.

To use Tor, a client builds circuits, each passing through an
\emph{entry}, a \emph{middle}, and an \emph{exit} relay. Each circuit supports
multiplexing multiple \emph{streams} of end-to-end TCP communication with
internet services. When a new TCP connection to a service is requested by an
application (e.g., Tor Browser), the Tor client will use fixed-size
application-layer control messages called \emph{cells} to instruct the exit
relay to (1)\;resolve the service's domain name, and (2)\;make a TCP connection
to the service. Each of a circuit's TCP byte-streams is subsequently forwarded
bidirectionally through the circuit in data cells. Circuit traffic observed
from a single network location can be represented as a time-ordered
sequence of (direction, time) pairs (one for each cell sent through a circuit),
which we call a \emph{cell trace}.

A TCP stream may be assigned to any circuit with an exit relay that
allows connection to the destination's IP address and port; if no such circuit
exists, a new one is built after choosing an exit independently at random
from among those with conforming exit policies and weighted by
relay bandwidth to balance load.
However, Tor Browser employs additional stream assignment rules.
When loading a webpage URL, Tor Browser computes the URL's first-party domain
name (FPDN) and instructs the Tor client to assign all streams created to load
that URL (including those to third-party domains to load embedded objects) on
a circuit uniquely associated with the FPDN and isolated from other streams.

Browsing to a page of a new website in Tor Browser will result in a unique FPDN
and a new circuit tasked with first resolving a DNS query for the FPDN and then
loading the page, while subsequent subpages of that website will
be loaded through the same circuit. \citeauthor*{USENIX:CheJanTro22} thus
recognized that (1)\;the FPDN in the circuit's first DNS query can be used to
\emph{label} the website of a circuit's cell trace, and (2)\;an adversary running exit relays
can observe \emph{genuine} cell traces and their domain name labels, which can be used to
train WF classifiers and produce more realistic estimates of WF
performance~\autocite{USENIX:CheJanTro22}. (Non-exit relays can observe cell
traces but not domain names due to onion routing~\autocite{SP:SyvGolRee97}.)
Unfortunately, their study considered a fully online setting to avoid
persistently storing genuine data or trained WF classifiers; thus a new
measurement is needed to build on the methodology.

In WF, an adversary uses the volume and timing of traffic to
infer which website is being visited~\cite{CCS:JAADG14}. This attack is suited
to breaking the privacy of traffic sent through VPNs or Tor because their traffic
plaintext and destination are unobservable. In a WF attack, the adversary observes
the client and its traffic.
Typical WF attacks use machine-learning classifiers trained on traffic
traces labeled with the destination website (e.g.,
\cite{CCS:SIJW18,USENIX:SJGLZX23,deng2025countmamba}).
The classifiers are applied to traffic traces from the target client to identify
the destination website. In the Tor setting, unlike for VPN traffic,
WF classifiers are typically only given when a cell appears and in which direction because
of Tor's fixed-size cells, which can either be recorded directly by a malicious Tor relay or
reconstructed from TCP packet payloads by a network observer.

\subsection{Measurement Process}

We designed a measurement process that employs one or more Tor exit relays to
safely measure genuine Tor cell traces and FPDN labels.
The traces and labels are collected into a \emph{dataset} for subsequent analysis.
Each participating relay runs a patched version of Tor
that we modified to support our measurement as follows.

\subsubsection{Circuit Selection}
When a relay observes a new circuit, it rejects any non-exit type circuit
(i.e.,\;onion-service and internal circuits) from measurement. Additionally, the
relay applies a probabilistic sampling procedure such that 80\% of exit-type
circuits are rejected during \emph{high-volume} measurement intervals, and 98\%
of exit-type circuits are rejected during \emph{low-volume} measurement
intervals. Sampling helps us limit the total amount of data collected and provides
plausible deniability: any individual circuit created through a participating
relay is unlikely to exist in the dataset. Non-rejected circuits are selected
for further measurement.

\subsubsection{Circuit Measurement}
A relay internally stores circuit metadata and cell traces during
operation for the randomly selected exit-type circuits. We use an encoding function $H(x)$
to protect some of this metadata:
\begin{equation}\label{hashencode}
    H(x) = \b64(\sha(x || \salt{}))
\end{equation}
where \emph{salt} is chosen uniformly at random, fixed on all measurement relays for the duration of
the measurement period, and then destroyed. Tor's existing internal implementations are used for the
$\b64(\cdot)$ and $\sha(\cdot)$ functions. The relay iteratively constructs a
circuit metadata record for each selected circuit (applying $H(\cdot)$ to domain
names) until either the circuit closes or $N$ cells have been observed,
whichever occurs first.\footnote{We use $N=\num{5000}$ cells to remain
consistent with previous work.} The metadata record is then exported via Tor's
control interface to an external process that compresses it, encrypts it with a
public-key encryption scheme,\footnote{We encrypt to an offline secret key to
prevent on-device decryption.} and writes it to persistent storage.

Each metadata record includes the following as exemplified in~\autoref{circmeta}:
(1)\;\emph{day}: an integer number of
days that have elapsed since the start of the measurement; (2)\;\emph{domain}:
$H(d)$ where $d$ is the domain name of the circuit's first exit
stream;\footnote{Circuits for which the first exit stream connects to the
destination with an IP address instead of a domain name are rejected from
measurement.} (3)\;\emph{shortest\_private\_suffix}: $H(s)$ where $s$ is the
shortest private suffix of the pre-image of \emph{domain} computed using
Mozilla's public suffix list and
\texttt{libpsl}~\autocite{libpsl};
(4)\;\emph{port}: the server port used when connecting the circuit's first exit stream to its
destination; and (5)\;\emph{cells}: a list of at most $N$ cell metadata items.
Each cell metadata item is a 4-tuple containing the time the cell was observed
relative to the circuit's creation time, an integer encoding the cell's direction,
and two integers encoding the cell and relay command, respectively~\cite{DinMat2004}.

\begin{figure}[t]
\begin{lstlisting}[
	basicstyle=\scriptsize,
	label=circmeta,
	caption={Example circuit metadata record.},
]
{
  "day":2,
  "domain":Dnqty37vYTIEivWhAEikb7HlJOzWXEZ2Rw05iicG7e8,
  "shortest_private_suffix":bIKFK8gYicwptEMM1Goxlo7KredMMFx48VD0MpXn9zc,
  "port":443,
  "cells":[
    [ 0.000015, 1,10,0],//client->exit: create
    [ 0.000463,-1,11,0],//exit->client: created
    [10.932340, 1, 9,1],//client->exit: relay_early.begin
    [12.070954,-1, 3,3],//exit->client: relay.connected
    [13.421017, 1, 9,2],//client->exit: relay_early.data
    [13.421030,-1, 3,2],//exit->client: relay.data
  ]
}
\end{lstlisting}
\end{figure}

\subsection{Safety and Ethical Considerations}
Our measurement process was designed to prioritize safety and
privacy and was developed through consultation with our organization's
Institutional Review Board (IRB) and the Tor Research Safety Board
(TRSB)~\autocite{trsb}.
Our IRB qualified our study as non-human-subject research, and our TRSB
interaction resulted in a ``No Objections'' score and instructions to move
forward with our plans. Extended analysis of the safety, risks, and benefits of
our measurement and details about IRB and TRSB interactions are
provided in~Appendix~\ref{app:ethics}.
 \section{Measurement and Analysis}
\label{sec:measurement}

\subsection{Measurement Details}

We execute a large-scale Tor measurement study following our methodology
from~\autoref{sec:methodology}.
First, we run a total of eight exit relays, four on each of two identical
machines hosted by the Calyx Institute (a nonprofit research and education
organization located in NY, USA). Each machine is equipped with 2 Intel Xeon
E5-2695 v2 12-core CPUs (48 hyper-threads in total) and connected to an
unmetered 1 Gbit/s symmetric network access link.
Second, we run a measurement over a 13~week period in 2023; we assign weeks 1,
7, and 13 as high-volume intervals, and the remaining 10 weeks as low-volume
intervals. We combined all recorded circuit metadata records into a single
dataset which we call \dataset\footnote{GTT: an acronym for ``Genuine Tor
Traces''; 2023: the year of measurement.}~\citeanon{gtt23_10620520}.

\subsection{Data Analysis}
\label{sec:measure:dataset}

 In
total, \dataset{} contains \num{13900621} circuits, \num{10557898} of which were
observed during the high-volume weeks (1, 7, and 13) and \num{3342723} of which
were observed during the remaining 10 low-volume weeks.

\begin{figure}
	\centering
	\includegraphics[width=0.99\columnwidth]{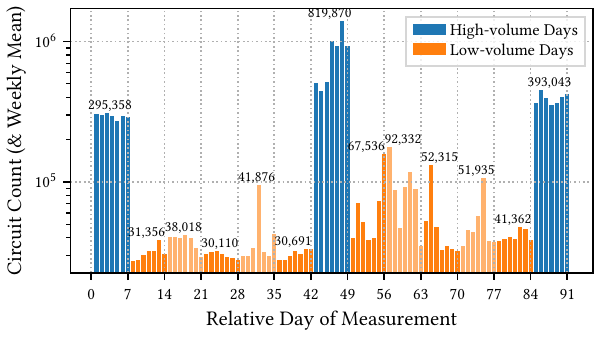}
	\caption{
		The daily total (bars) and weekly mean (text) number of circuits
    during our 13 week measurement.
	}
	\label{fig:measure:circs_per_day}
\end{figure}

\looseness-1
The daily total and weekly mean number of \dataset{} circuits are shown in
\autoref{fig:measure:circs_per_day}; the daily mean during high-volume weeks is
\num{502757} and the daily mean during low-volume weeks is \num{47753}. We
observe a slight increase in circuit counts during the latter half of the
measurement period which we attribute to natural fluctuation in network usage
and the load-balancing weights used for relay selection.

\begin{figure}
	\centering
	\includegraphics[width=0.99\columnwidth]{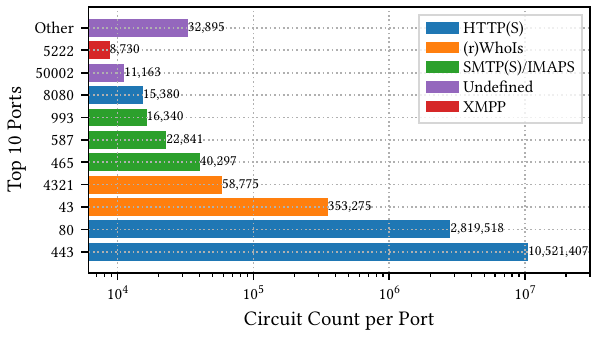}
	\caption{
		The total number of \dataset{} circuits by server port,
		with IANA-assigned service names~\autocite{IANA2023}.
}
	\label{fig:measure:circs_per_port}
\end{figure}

\dataset{} contains circuits measured across \num{68} unique destination server ports.
The distribution of the number of measured circuits across
the top-ten most-popular service ports is shown in
\autoref{fig:measure:circs_per_port} (with a logarithmic x-axis, and the
IANA-assigned service names shown in the legend). We observe that \num{13356305}
circuits (96\%) use ports 80, 8080, or 443 to connect their first stream to a
destination service; these ports are assigned to HTTP and HTTPS by the
IANA~\autocite{IANA2023}. The vast majority of the remaining circuits use port 43 or
4321, which are respectively assigned to WhoIs and Remote WhoIs services by the IANA. 
Frequent connections to these ports have been observed in
prior studies of Tor exit
traffic~\autocite{sonntag2017traffic,sonntag2019malicious}:
Sonntag observed that they
corresponded to a large number of reverse DNS lookups scanning several large
networks~\autocite{sonntag2019malicious}.

\begin{figure}
	\centering
	\includegraphics[width=0.99\columnwidth]{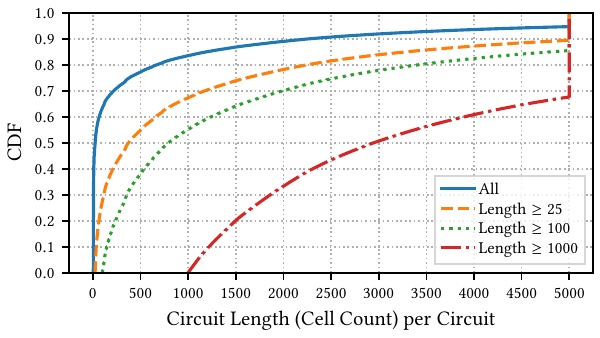}
	\caption{
		Cumulative distribution of the number of cells per circuit over
		subsets of \dataset{} circuits.
	}
	\label{fig:measure:cells_per_circ}
\end{figure}

The cumulative distribution of the number of observed cells per \dataset{} circuit
is shown in \autoref{fig:measure:cells_per_circ}. We were surprised to find that
most circuits are extremely short: the median number of cells over all circuits
is just 25, which would support at most
\num{10.5}\;KB of application payload
after accounting for control cells and cell-header overhead. 
For comparison, we also plot in \autoref{fig:measure:cells_per_circ} the circuit
length distribution for the subsets of circuits containing at least 25, 100, and
\num{1000} cells, respectively corresponding to \num{10.5}, \num{47.8}, and
\num{496}\;KB of application payload.
For reference, the HTTP Archive reports that over 90\% of webpages have a
transfer size greater than 450\;KB across samples of 12 and 16 million desktop
and mobile URLs, respectively (see \autoref{tab:http_size}). Thus, we believe
that most \dataset{} circuits did not carry full webpage transfers.

\begin{table}
    \centering
    \small
    \begin{threeparttable}
    \caption{Total Web Page Transfer Size (KB) in 2023-08$^{\dagger}$}
    \label{tab:http_size}
    \begin{tabularx}{\columnwidth}{X | r
      S[table-format=3]
      S[table-format=3]
      S[table-format=3]
      S[table-format=3]
      S[table-format=3]
    }
        \toprule
        {\textbf{Client}} & {\bf Percentile:} & {\textbf{10th}} & {\textbf{25th}}
                          & {\textbf{50th}} & {\textbf{75th}} & {\textbf{90th}} \\
        \midrule
        Desktop && 547 & 1226 & 2484 & 4967 & 9744 \\
        Mobile && 457 & 1063 & 2179 & 4377 & 8946 \\
        \bottomrule
    \end{tabularx}
    \begin{tablenotes}
        \item[$\dagger$] HTTP Archive: \href{https://httparchive.org/reports/state-of-the-web}{https://httparchive.org/reports/state-of-the-web}
    \end{tablenotes}
    \end{threeparttable}
\end{table}

\begin{figure}[t]
	\centering
	\includegraphics[width=0.99\columnwidth]{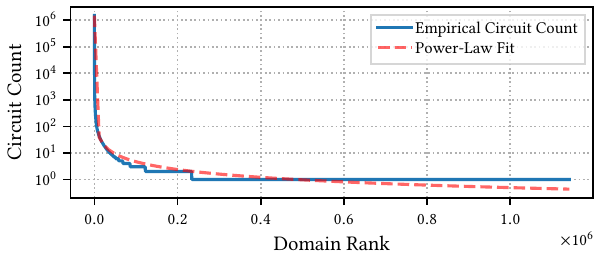}
	\caption{
		The number of \dataset{} circuits per domain; we observe a close fit to a
		power-law distribution.
	}
	\label{fig:measure:domain_circ_count}
\end{figure}

\dataset{} contains circuits measured across \num{1142115} unique destination domains.
The distribution of the number of measured circuits per
domain is plotted in \autoref{fig:measure:domain_circ_count}. We observe a close
fit to a power-law distribution (shape=\num{0.023}, loc=\num{0.769},
scale=\num{1495234}), where few popular domains dominate the measurement while a
long tail exists with just a single circuit measured for \num{908422} (80\%) of
the domains.

Note that obtaining realistic base rates for the domains visited by Tor users is a major
advantage of \dataset{} over synthetic datasets. In open-world binary
classification, the negative class is composed of traces to all sites other than
the monitored ones. Thus, the false-positive rate, which is crucial for
estimating precision~\cite{SP:Wang20}, depends on the base rates in the negative
class.
Similarly, in a multiclass setting (open or closed world), overall WF accuracy
depends on the base rates of each class.

\begin{figure}
	\centering
	\includegraphics[width=0.99\columnwidth]{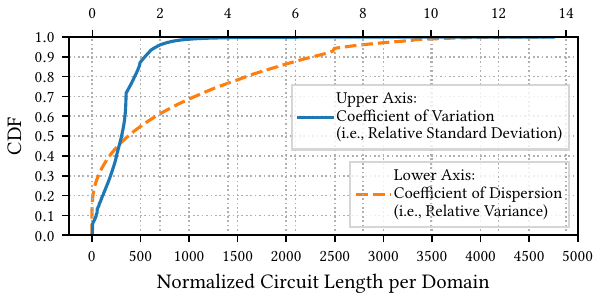}
	\caption{
		Cumulative distribution of circuit length variation across
		domains with at least two \dataset{} circuits.
	}
	\label{fig:measure:domain_circ_length}
\end{figure}

\autoref{fig:measure:domain_circ_length} shows the cumulative distribution of
two measures of circuit length variability for each domain with more than one
\dataset{} circuit. The median Coefficient of Variation (i.e., the standard
deviation divided by the mean) shows that more than half
of the domains have a circuit length standard deviation greater than the mean,
while the Coefficient of Dispersion (i.e.,\;the variance
divided by the mean) shows that most domains have a relative
variance in circuit lengths of multiple hundreds of cells. The high variability
in circuit lengths is consistent with our prior observation that most of the
measurement circuits are short, and suggests that many Tor circuits may
completely or prematurely fail.

 \section{Evaluation}
\label{sec:evaluation}

In this section, we compare \dataset{} and synthetic datasets to understand how
well the synthetic datasets model some of the genuine data characteristics that
are important for WF.

\subsection{Deficiencies of Synthetic Datasets}
\label{sec:analyze:datasets}

We survey 28 datasets proposed for WF tasks covering the years 2008--2025
In~\autoref{tab:selectdatasets} we provide an overview of the
properties of a subset of the surveyed datasets, selected for their size,
complexity, and frequency with which they are used to evaluate later attacks.
See \autoref{tab:relwork-data} in Appendix~\ref{app:datasets} for the full comparison.

Like \dataset{}, these datasets also consist of Tor traffic traces labeled with
a destination domain, and they record traffic that actually transited the Tor
network and connected to some third-party server. However,
we find that every dataset exhibited
similar deficiencies:
(1)\;they consist of only web traffic;
(2)\;they are collected using simplistic user models and static software tools,
almost exclusively at the client position;
(3)\;they primarily focus on fetching popular webpages; and
(4)\;they do not contain informed base rates.
In contrast, real Tor clients use a wide variety of software and software
versions, interact with non-web services, and do more than just
non-interactively fetch selected webpages.
These deficiencies make it difficult to
use existing datasets to draw meaningful conclusions about the effectiveness of
a WF attack directed at real Tor users~\autocite{CCS:JAADG14,USENIX:CheJanTro22}.

In comparison, \dataset{} is the only dataset with traces sampled from genuine
traffic created by real Tor users interacting with real internet services at
natural base rates. \dataset{} is not limited to only web traffic: it contains
traces of different types of internet activity and supports the evaluation of WF attacks
and defenses based on websites' first-party domain names (see~\autoref{sec:methodology:background}).
These traces better represent
the WF problem, where an adversary observes undifferentiated traffic from real users and cannot
assume that the traffic is just to index web pages or is even to a website at all. Thus,
\dataset{} can serve to more accurately evaluate the
threat posed by a real-world WF adversary. Moreover, \dataset{} is larger
than the previous largest dataset (\citeauthor*{NDSS:RPJvJ18}'s AWF dataset
\cite{NDSS:RPJvJ18}) by an order of magnitude which is important to assess
modern deep learning attacks requiring many training examples. Extended results
and analysis from our dataset survey appear in
Appendix~\ref{app:datasets}.

{
\makeatletter
\def\TY@box@v#1{$\vcenter \@startpbox{\csname TY@F\the\TY@count\endcsname}#1\arraybackslash\tyformat
                              \insert@column\@endpbox$}
\makeatother

\begin{table}
  \setlength{\tabcolsep}{2.5pt}
  \centering
  \footnotesize
  \begin{threeparttable}
  \caption{Select WF Datasets (full details in~\autoref{tab:relwork-data})}
  \label{tab:selectdatasets}
  \begin{tabulary}{\columnwidth}
  {
    llS[table-format=1.1e1, tight-spacing=true]L
  }
      \toprule
      {\textbf{Dataset}} & {\textbf{Year}} & {\textbf{Size}} & {\textbf{Description$^{\dagger}$}} \\
      \midrule
      $k$-NN~\autocite{USENIX:WCNJG14} & 2014 & 1.4e4 & Web, top index pages \\
      \rowcolor{gray!10} AWF~$\mathit{CW}_{900}$~\autocite{NDSS:RPJvJ18} & 2017 & 2.3e6 & Web, top index pages \\
      AWF Open~\autocite{NDSS:RPJvJ18} & 2017 & 8e5 & Web, top index pages \\
      \rowcolor{gray!10} DF~\autocite{CCS:SIJW18} & 2018 & 1.4e5 & Web, top index pages \\
      GoodEnough~\autocite{pulls2020towards} & 2020 & 2e4 & Web, top index pages + subpages \\
      \rowcolor{gray!10} BigEnough~\autocite{SP:MHORHW23} & 2021 & 3.8e4 & Web, top index pages + subpages \\
      Multi-tab~\autocite{SP:DYLZLXXW23} & 2022 & 5.7e5 & Web, top index pages, multiple tabs \\
      \rowcolor{green!10} \dataset{} & 2023 & 1.4e7 & Genuine traffic, real user behavior, visited services, natural base rates \\
      \bottomrule
  \end{tabulary}
  \begin{tablenotes}
      \item[$\dagger$] All but \dataset{} synthetically fetch webpages using automated tools.
  \end{tablenotes}
  \end{threeparttable}
\end{table}
}

\begin{figure*}
	\centering
  \begin{subfigure}[t]{0.32\textwidth}
    \includegraphics[width=\textwidth]{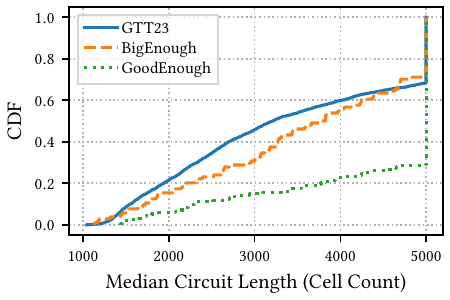}
  \end{subfigure}
  \begin{subfigure}[t]{0.32\textwidth}
    \includegraphics[width=\textwidth]{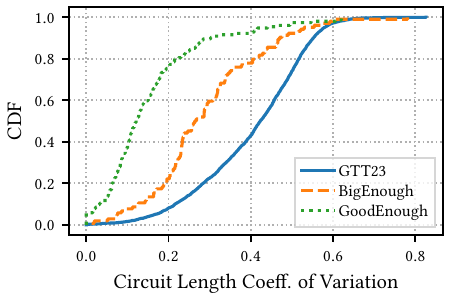}
  \end{subfigure}
  \begin{subfigure}[t]{0.32\textwidth}
    \includegraphics[width=\textwidth]{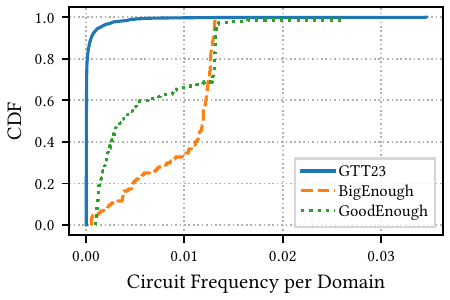}
  \end{subfigure}
	\caption{
    Per-domain statistics computed from the \dataset{},
    BigEnough~\autocite{SP:MHORHW23}, and GoodEnough~\autocite{pulls2020towards}
    datasets. For statistical rigor, here we consider traces with at least
    \num{1000} cells and, among those, domains with at least 30 traces.
	}
	\label{fig:stats:genuine-v-synthetic}
\end{figure*}

\subsection{Genuine and Synthetic Disparities}

We analyze the statistical disparities between \dataset{} and synthetic
datasets to understand dataset quality. We place particular emphasis on
the trace features found in prior work~\autocite{USENIX:HayDan16} to be
informative for WF. We focus our analysis on two recent synthetic datasets,
BigEnough~\autocite{SP:MHORHW23} and GoodEnough~\autocite{pulls2020towards},
that were specifically designed to model
web\emph{site} fingerprinting; both datasets contain at least ten
pages per website and so they represent among the highest website diversity
of the datasets surveyed.

\subsubsection{Dataset Composition}
The \dataset{} dataset contains traces generated from real users interacting
with any services accessible via the internet (including non-web services),
whereas synthetic datasets such as BigEnough and GoodEnough contain traces
generated from automated visits to small number of popular websites.
Empirical data from this work and previous work suggests that Tor users do
not just visit popular websites. First, \autoref{fig:measure:circs_per_port}
shows that a long tail ($\approx$4\%) of \dataset{} traces are generated from
interactions with hosts not running on known web ports such as 80, 443, or 8080.
Second, a privacy-preserving measurement of the Tor network performed in
2018~\autocite{IMC:MWJJS+2018} determined that over~20\% of web streams exiting
the Tor network access a host not in the Alexa~Top~1~Million list. This
long tail of activity is not reflected in the synthetic datasets and is likely to make the
WF classification task more
difficult~\autocite{NDSS:PLPEZH16}.

\subsubsection{Data Modeling}
Even simple features computed from the synthetic datasets do not accurately
model genuine Tor traces. Consider, for example, overall trace length, a feature
shown in prior work to be informative in the WF task~\autocite{USENIX:HayDan16}.
The leftmost plot in \autoref{fig:stats:genuine-v-synthetic} shows, for all 3
datasets, the distribution of each domain's median circuit length (cell count)
for circuit traces with at least \num{1000} cells and among those, domains with at least 30
traces. The plot
shows that \dataset{} traces tend to be shorter than synthetic dataset traces.
GoodEnough traces, in particular, tend to be much longer than genuine traces:
roughly 70\% of GoodEnough domains have a median circuit length of \num{5000}
cells (the capture limit), whereas this is true of only 32\% of
domains in the \dataset{}~dataset. Inaccurate data modeling makes it difficult
to draw meaningful conclusions from the synthetic
datasets~\autocite{USENIX:AQPWPWCR22}.

\subsubsection{Intra-class Variance}
Genuine user traces contain a much richer set of activity than is generated by
synthetic, automated crawls to webpages. Genuine traces may be generated from
various unpredictable user-initiated behaviors and processes
and may reflect complex,
interactive sessions with internet hosts, whereas synthetic traces are usually
generated by a single, fixed crawling application such as tor-browser-selenium~\autocite{tor-browser-selenium}
and are limited to simple page accesses. The center plot in
\autoref{fig:stats:genuine-v-synthetic} shows the distribution of the
coefficient of variation of trace length---that is, the ratio of the trace length's
standard deviation to the mean---across the domains of each dataset. At nearly every
percentile, the coefficient is higher for \dataset{} domains than it is for
BigEnough and GoodEnough domains, suggesting that \dataset{} traces exhibit
higher variation. The higher variance for each domain suggests that WF is more difficult
on genuine than on synthetic traces.

\subsubsection{Base Rates}
Recall that the frequency
of website occurrence in the \dataset{}~dataset is characterized by a few heavy
hitters and a long tail of rarely accessed sites (see \autoref{fig:measure:domain_circ_count}).
In contrast, the rightmost
plot of \autoref{fig:stats:genuine-v-synthetic} shows that most domains in the
synthetic datasets occur with much higher frequency. For example, the
median domain occurs with frequency \num{5e-4} in \dataset{}. In comparison,
the median domain in the BigEnough and GoodEnough dataset occur with frequencies
that are orders-of-magnitude greater, \num{1e-2} and \num{4e-3}. Base rate
realism is an important aspect of evaluating WF~attacks because increasingly low
false positive rates are needed to maintain precision at low base rates of
occurrence~\autocite{CCS:JAADG14, CCS:CNWJG14, SP:Wang20}.
Precisely fingerprinting most websites in
\dataset{} requires orders-of-magnitude
lower false positives rates compared to BigEnough and GoodEnough.

 \section{Conclusion and Future Work}
\label{sec:conclusion}

Our goal in this work is to ground the future study of realistic WF attacks and
defenses while promoting reproducible research. Toward this goal,
we designed a safe Tor measurement methodology through consultation with the
TRSB~\cite{trsb} and our institution's IRB. We executed the methodology across
eight Tor exit relays to measure \num{13900621} circuits and compose \dataset{},
the first dataset of genuine Tor traces. We analyze \dataset{} to better
understand the characteristics of genuine Tor data, and evaluate the extent to
which existing synthetic WF datasets misrepresent genuine characteristics.
We conclude that, because \dataset{} contains genuine traces of websites
accessed by real Tor users at completely natural base rates, it is more
realistic than any existing synthetic dataset and thus it will enable WF evaluations
that more accurately estimate real-world WF performance.

\dataset{} has been available to researchers upon request since 2004
since~\cite{gtt23_10620520}, and it has already been used to improve
our understanding of WF on Tor.
Jansen et al.~\cite{jansen2023repositioning} develop the Retracer methodology for performing
WF analysis on Tor trace datasets that, like \dataset{}, are collected at the exit relay.
Retracer modifies the traffic traces to appear more as they would to an adversary
observing the client. The results indicate that a WF adversary is likely to obtain much lower
accuracy than synthetic datasets have indicated. Jansen~\cite{cellshift-ndss2026}
further develops this methodology. Similarly, Deng et al.~\cite{deng2025beyond} use GTT23 to
perform a WF analysis where the adversary is detecting connections to a monitored set of sites.
Their results indicate that the Var-CNN classifier obtains higher accuracy than the DF classifier
used by Jansen et al.

\dataset{} can be used to conduct both training and testing of WF classifiers in
either an offline or an online streaming model. During training, \dataset{} can
provide genuine examples not only of a target class of websites, but also of the
background class; negative class examples are important for training classifiers
to be robust against the high-diversity WF patterns we find in practice.
However, it is even more important to incorporate genuine traces while testing
WF classifiers to accurately represent the challenge facing a real-world
adversary deploying an attack. Future work should consider new WF classification
techniques that are specifically developed and tuned to genuine traces to
further improve our understanding of the real world WF performance an adversary
can expect to attain.

One limitation of \dataset{} is that the labels were created by randomizing
website domains with an irreversible hash function due to privacy and ethical
requirements. While this may seem to inhibit our ability to synthesize
\dataset{} with other labeled datasets, many emerging self-supervised ML
training methods~\autocite{PoPETS:OMRWH21, CCS:BaBoHo23, CCS:SMRW19} have
unsupervised pre-training phases that learn how to generally differentiate
websites without the need for labels. Such approaches could leverage \dataset{}
for pre-training to benefit from the genuine patterns it contains, and then use
a different labeled dataset with many fewer traces for fine-tuning the
pre-trained models.

There are also applications outside of WF. The traffic statistics of \dataset{}
regarding Tor usage are of independent interest, since previous Tor measurement
studies are already outdated by more than five
years~\autocite{IMC:MWJJS+2018,CCS:JanTraHop18}. Additionally, \dataset{} could
be used to enhance the realism and fidelity of Tor network testbeds and
simulations~\autocite{NDSS:JanHop12,ATC:JanNewWai} since the traces encode
the timing and patterns of genuine user activities. \section*{Acknowledgments}
This work was supported by the Office of Naval Research (ONR). 

{
\printbibliography
}

\appendix
\section*{Appendices}

\section{Ethics}
\label{app:ethics}

In this section we describe our consideration of ethics as we developed
and executed our Tor network measurement plan.

\subsection{Safety}
The primary safety guidelines driving our measurement study include collecting
only what is adequate, relevant, and necessary for our purpose (data
minimization), limiting the granularity of the data that is collected, ensuring
security of the measurement infrastructure and confidentiality of the
measurement data, and limiting access to measurement infrastructure to a single
author.

We took numerous specific precautions to make sure our measurement is as safe as
possible. First, our measurement is carried out exclusively by relays under our
control, which are identified as part of the same relay family following relay
operator best practices. Our measurement is completely passive: it does not
alter the default Tor protocol and does not add any load to the network.
Second, we use probabilistic sampling to randomly select circuits for
measurement. While sampling enables us to limit the total data volume and
maintain an accurate, genuine distribution of website access frequency, it also
provides plausible deniability for Tor users because any individual circuit that used our
relays during measurement is still unlikely to appear in the dataset.
Third, we chose to exclude absolute timestamps from the dataset, and we do not
precisely specify when the measurement occurred. Precise absolute timing
information is not required for WF and there may be some privacy benefit to
excluding it.
Fourth, we use an irreversible encoding function (\autoref{hashencode}) to
protect the destination domain name and provide a website label that is
consistent within our dataset but whose mapping can no longer be reproduced
since the salt was erased.
Finally, we encrypt all records using the public part of a public-key encryption
scheme before writing to persistent storage or transferring over a network.
The secret key is stored offline to prevent unauthorized decryption.

\subsection{Risks}
Our measurement process is designed to be consistent with Tor's threat model
in which exit relays can only observe the destination side of the circuit but
cannot link this data to the client side of the circuit. Our measurement data
alone cannot be used to identify any individual client.

An adversary could possibly augment our exit-side data patterns with their own
client-side flows collected during our measurement period and attempt an
end-to-end correlation attack to link a client they observed to a record in our
dataset. We mitigated this risk by mapping the true domain name to a random
string label using an encoding function that cannot be reproduced or reversed
(\autoref{hashencode}), so that a successful correlation attack would still be
unable to precisely identify the visited destination. In order to fully
deanonymize a flow, the adversary would need to combine the correlation attack
with a confirmation attack wherein they link a known \emph{labeled} flow to a
record in our dataset, enabling them to map the random string output of
\autoref{hashencode} to a known label. While not impossible, we believe that
such a multi-step attack would be imprecise and thus there is a low risk that
it would be successful on a large scale in practice.

\subsection{Benefits}
Measurement and publication of our genuine website fingerprinting dataset can
help researchers and practitioners improve Tor's protections against traffic
analysis attacks and better quantify the risks that traffic analysis poses to
users. Our dataset improves realism relative to synthetic datasets and can help
other researchers quickly get started designing WF attacks or defenses without
first needing to perform possibly network-invasive measurement themselves. Our
dataset also promotes reusability and reproducible research, and may make it
more likely that WF defenses developed in research will work well in the real
world.

\subsection{Alternatives}
One alternative to our measurement approach is to recruit volunteers to opt-in
to a measurement study and then only collect patterns on circuits initiated from
volunteers. This alternative was rejected because it produces biased samples of
traffic at synthetic base rates and it involves human subjects which
would add significant additional risk to our study.

\subsection{Board Review}
We consulted our organization's Institutional Review Board (IRB) to determine if
our measurement study required further evaluation as human-subject research. The
board determined that our study qualifies as non-human-subject research because
we are not directly interacting with subjects nor will the network data we are
gathering enable us to identify the people who may be using the network.

We also submitted a document detailing our measurement plans and ethical
analysis to the Tor Research Safety Board
(TRSB)
for review. The review of our initial submission identified several important
aspects of our measurement plan that were underspecified, including details
around the scope of measurement and our plan for informing stakeholders.
Additionally, our submission presented multiple possible methods of dataset
release with a request for feedback on the most appropriate. The board expressed
preference for a controlled release where the dataset would only be shared with
verified researchers upon request. Following an interactive discussion, we
extended our measurement plan to include the requested details and plans for a
controlled release of the dataset.
Following re-review of our updated submission, we received a ``No Objections''
score and were instructed to move forward with our plans.

\subsection{Community Notification}
We notified the Tor community of our plans through a post to the tor-project mailing
list~\citeanon{Jan2023}.
The primary feedback we received was that we should collect \emph{more} data,
which we decided against following the principle of data minimization.

 \section{Survey of Existing WF Datasets}
\label{app:datasets}

We surveyed prior work related to website fingerprinting attacks in order to
better understand the datasets used to quantify attack effectiveness. We
evaluated each dataset among a number of different dimensions, as follows.
\begin{description}
  \item[Year:] the time the dataset was collected;
  \item[Activity:] the kind of user behavior contained in the dataset;
  \item[User model:] the way in which users perform the activity;
  \item[Trace generation software:] the tools used in activity creation;
  \item[Size:] the number of classes and traces in the dataset;
  \item[Availability:] the accessibility of the dataset to others;
  \item[Attacks:] the WF attacks originally evaluated on the dataset.
\end{description}
We also noted how each dataset was recorded (that is, the software used and
trace observation point, if provided).

The summary of results is shown in \autoref{tab:relwork-data}. All datasets
surveyed were composed of primarily web activity. Most datasets assume users
interact with popular websites, usually those present in the now-discontinued
``Alexa Internet'' top websites ranking. A few works consider more sophisticated
user behaviors:
\citeauthor*{Herrmann2009}~\cite{Herrmann2009} collect URLs obtained from
monitoring an academic proxy server they had access to;
\citeauthor*{CCS:JAADG14}~\cite{CCS:JAADG14} collect URLs obtained from
volunteers browsing the Internet;
\citeauthor*{NDSS:PLPEZH16}~\cite{NDSS:PLPEZH16} considered URLs obtained
from observing Tor HTTP exit traffic, as well as from
interacting with popular Internet services such as Twitter and Google;
and \citeauthor*{SP:DYLZLXXW23}~\cite{SP:DYLZLXXW23} collected URLs from
volunteers browsing the Internet.

The task designated for each dataset may vary. For example,
RND-WWW~\cite{NDSS:PLPEZH16}, \citeauthor*{CCS:JAADG14}~\cite{CCS:JAADG14},
GDLF-25~\cite{PoPETS:OMRWH21}, GoodEnough~\cite{pulls2020towards},
BigEnough~\cite{SP:MHORHW23}, ALEXA-WSC-FG/BG~\cite{mitseva2024stop}, and
CW/OW~\cite{zhao2024towards} are designed to incorporate multiple pages
for each of many websites. AWF~Recollect~\cite{NDSS:RPJvJ18} and
WTT-Time~\cite{PoPETS:OhSunHop19} are designed to explore aspects of concept
drift. $\mathit{DS_{Tor}}$ contains .onion sites in addition to ordinary
websites. Multi-tab~\cite{SP:DYLZLXXW23,zhao2024towards} contains only browsing
behavior occurring simultaneously in several browser tabs.

All extant datasets are collected synthetically with an automated crawl, often using a
single set of software to generate flows
(\citeauthor*{CCS:JAADG14}~\cite{CCS:JAADG14} and
\citeauthor*{SP:DYLZLXXW23}~\cite{SP:DYLZLXXW23} both consider the effect that
varying versions of Tor Browser Bundle (TBB) may have on attacks). Additionally,
nearly every work
uses {\tt tcpdump} to collect packet traces on the \emph{client} generation machine. Only
GoodEnough, BigEnough, and $D(\mathrm{tbs, tor})$~\cite{PoPETS:JanWai23}
collect cell traces using the {\tt tor} process directly; GoodEnough and
BigEnough are collected at the client position, whereas 
$D(\mathrm{tbs, tor})$ is collected at the guard position.

Inconsistent purposes, over-simplified user models, and static collection
software make it difficult to draw meaningful conclusions about the
effectiveness of a WF attack directed at real Tor users. Real Tor clients use a
wide variety of software (most network applications supporting SOCKS5 can be
used with Tor), interact with non-web services, and do more than just non-interactively fetch
random pages on the web. In contrast, \emph{GTT23 is the only dataset addressing
these weaknesses}---it contains traces from real Tor client interacting with real
internet services. Moreover, GTT23 is larger than the previous largest dataset
by an order of magnitude (AWF $\mathit{CW_{900}}$~\cite{NDSS:RPJvJ18}) and is
larger than most other existing datasets by multiples orders of magnitude; this
volume of data is important when training modern deep learning models which may
require millions of examples to be effective.

\newgeometry{left=0.15in,right=0.15in,top=1in,bottom=1in}
{
  \onecolumn

  \makeatletter
  \def\TY@box@v#1{$\vcenter \@startpbox{\csname TY@F\the\TY@count\endcsname}#1\arraybackslash\tyformat
                                \insert@column\@endpbox$}
  \makeatother

  \begin{landscape}
    \begin{table}[h]
      \setlength{\tabcolsep}{2.5pt}
      \centering
      \footnotesize
      \caption{Summary of website fingerprinting datasets curated over the past
      15~years. The `$\bot$' symbol is used to indicate a dataset is unnamed, and 
      the `-' symbol is used when a cell's contents are identical to
    the above cell. When the year of data collection is not mentioned, we assume
  it is around (``ca.'') the associated article's publication date. Not all
  datasets describe their trace generation software with the same specificity.
  $\mathit{N, N_{C}, N_{I}, N_{Bg}}$ are the total number of traces in the
  dataset, the number of positive classes, the number of instances per positive
  class, and the number of background traces. The ``Attacks'' column shows a
  list of WF attack papers evaluated on the dataset.}
  \label{tab:relwork-data}

      \begin{tabulary}{618pt} {
        llrcLLL
        S[table-format=1.1e1, tight-spacing=true]
        S[table-format=4]
        S[table-format=4]
        S[table-format=1.1e1, tight-spacing=true]
        cC
        }
        \toprule
        {\bf Ref.} &
        {\bf Name} &
        {\bf Year} &
        {\bf Activity} &
        {\bf Activity Detailed} &
        {\bf User Model} &
        {\bf Trace Gen. Software} &
        {\bf $\bm{N}$} &
        {\bf $\bm{N_{\mathbf{C}}}$} &
        {\bf $\bm{N_{\mathbf{I}}}$} &
        {\bf $\bm{N_{\mathbf{Bg}}}$} &
        {\bf Available} &
        {\bf Attacks}
        \\ \midrule
\rowcolor{gray!10} \autocite{Herrmann2009} &
        $\bot$~(Hermann) &
        2008 &
        Web &
        Links from real-world academic proxy server &
        Index page &
        Autofox &
        8.5e3 &
        775 &
        \approx 10 &
        &
        \href{https://www-sec.uni-regensburg.de/website-fingerprinting/}
        {Dead link \faExternalLink} &
        \autocite{Herrmann2009} \\
\autocite{CCS:CZJJ12} &
        $\bot$~(Cai) &
        Ca. 2012 &
        Web &
        Alexa top sites &
        Index page &
        tor~0.2.1/2 &
        3.2e4 &
        800 &
        \approx 40 &
        &
        No &
        \autocite{CCS:CZJJ12} \\
\rowcolor{gray!10} \autocite{WPES:WanGol13} &
        levdata2 &
        Ca. 2013 &
        Web &
        Alexa top sites &
        Index page &
        tor~0.2.4.7; TBB~2.4.7 &
        4e3 &
        100 &
        40 &
        &
        \href{https://www.cs.sfu.ca/~taowang/wf/data/}
        {Online \faExternalLink} &
        \autocite{WPES:WanGol13, NDSS:PLPEZH16} \\
- &
        levdata3 &
        - &
        - &
        Popular blocked sites, Alexa top sites &
        - &
        - &
        9e2 &
        4 &
        10 &
        8.6e2 &
        - &
        - \\
\rowcolor{gray!10} \autocite{USENIX:WCNJG14} &
        $k$-NN &
        Ca. 2014 &
        Web &
        Sensitive sites, Alexa top sites &
        Index page &
        TBB~3.5.1; iMacros~8.6.0 &
        1.4e4 &
        100 &
        90 &
        5e3 &
        \href{https://www.cs.sfu.ca/~taowang/wf/data/}
        {Online \faExternalLink} &
        \autocite{USENIX:WCNJG14, WPES:WanGol13, PoPETS:WanGol16,
        PoPETS:OhSunHop19, CCS:SMRW19, NDSS:PLPEZH16, abe2016fingerprinting} \\
\autocite{CCS:JAADG14} &
        $\bot$~(Ju\'{a}rez)&
        Ca. 2014 &
        Web &
        Alexa top sites, volunteer browsing &
        Index page, visited pages &
        TBB~(2/3.X); Selenium &
        4.3e4 &
        200 &
        \approx 40 &
        3.5e4 &
        On request &
        \autocite{CCS:JAADG14} \\
\rowcolor{gray!10} \autocite{PoPETS:WanGol16} &
        $\bot$~(Wang) &
        2014 &
        Web &
        Sensitive sites, Alexa top sites &
        Index page &
        tor~0.3.6.4; TBB~3.6.4 &
        9e3 &
        100 &
        40 &
        5e3 &
        No &
        \autocite{PoPETS:WanGol16} \\
\autocite{NDSS:PLPEZH16} &
        RND-WWW &
        Ca. 2016 &
        Web &
        Twitter, Alexa one-click, Google Trends, Google Random, censored sites &
        Random subpage &
        TBB~3.6.1; Chickenfoot; iMacros; Scriptish &
        1.6e5 &
        1125 &
        40 &
        1.2e5 &
        \href{http://lorre.uni.lu/~andriy/zwiebelfreunde/}
        {Dead link \faExternalLink} &
        \autocite{NDSS:PLPEZH16, } \\
\rowcolor{gray!10} - &
        TOR-Exit &
        - &
        - &
        HTTP requests of real Tor users &
        Visited page &
        - &
        2.1e5 &
         &
         &
        2.1e5 &
        - &
        - \\
- &
        WEBSITES &
        - &
        - &
        Popular websites &
        Index page, random subpage &
        - &
        5.3e3 &
        50 &
        105 &
         &
        - &
        - \\
\rowcolor{gray!10} \cite{USENIX:HayDan16} &
        $\mathit{DS_{Tor}}$ &
        Ca. 2016 &
        Web &
        Alexa top sites, popular .onion sites &
        Index page &
        TBB; Selenium &
        1.1e5 &
        85 &
        \approx 90 &
        1e5 &
        \href{https://github.com/jhayes14/k-FP}
        {Dead link \faExternalLink} &
        \cite{USENIX:HayDan16, PoPETS:OhSunHop19} \\
\cite{NDSS:RPJvJ18} &
        AWF~$\mathit{CW}_{900}$ &
        2017 &
        Web &
        Alexa top sites &
        Index page &
        tor~0.2.8.11; TBB~6.5; Selenium &
        2.3e6 &
        900 &
        2500 &
         &
        \href{https://github.com/DistriNet/DLWF}
        {Online \faExternalLink} &
        \cite{NDSS:RPJvJ18, PoPETS:OhSunHop19, PoPETS:OMRWH21, CCS:SMRW19,
        PoPETS:BLKD19} \\
\rowcolor{gray!10} - &
        AWF Recollect &
        - &
        - &
        - &
        - &
        - &
        1e5 &
        200 &
        500 &
         &
        - &
        - \\
- &
        AWF Open &
        - &
        - &
        - &
        - &
        - &
        8e5 &
        200 &
        2000 &
        4e5 &
        - &
        - \\
\rowcolor{gray!10} \cite{CCS:SIJW18} &
        DF &
        Ca. 2018 &
        Web &
        Alexa top sites &
        Index page &
        tor-browser-selenium &
        1.4e5 &
        95 &
        1000 &
        4.1e4 &
        \href{https://github.com/deep-fingerprinting/df}
        {Online \faExternalLink} &
        \cite{CCS:SIJW18, PoPETS:OMRWH21, CCS:SMRW19, PoPETS:RSMGW20} \\
\cite{PoPETS:OhSunHop19} &
        WTT-time &
        2018 &
        Web &
        Alexa top sites &
        Index page &
        tor~0.4.0.8; tor-browser-crawler &
        8e4 &
        100 &
        300 &
        5e4 &
        On request &
        \cite{PoPETS:OhSunHop19} \\
\rowcolor{gray!10} \cite{pulls2020towards} &
        Good Enough &
        2020 &
        Web &
        Alexa top pages, random subpage &
        Index page &
        TBB~9.0.2 &
        2e4 &
        500 &
        20 &
        1e4 &
        \href{https://github.com/pylls/padding-machines-for-tor}
        {Online \faExternalLink} &
         \\
\cite{SP:Wang20} &
        $\bot$~(Wang) &
        2019 &
        Web &
        Alexa top sites &
        Index page &
        tor~0.4.0.1; TBB~8.5a7 &
        1e5 &
        100 &
        200 &
        8e4 &
        \href{https://github.com/literaltao/openwf}
        {Partially Online \faExternalLink} &
        \cite{SP:Wang20} \\
\rowcolor{gray!10} - &
        Wikipedia &
        - &
        - &
        Wikipedia browsing &
        Random subpage &
        - &
        2e4 &
        100 &
        100 &
        1e4 &
        - &
        - \\
        \cite{PoPETS:OMRWH21} &
        GDLF-25 &
        Ca. 2021 &
        Web &
        Alexa top sites &
        Random subpage &
        tor-browser-crawler &
        9.4e4 &
        2400 &
        39 &
         &
        On request &
        \cite{PoPETS:OMRWH21} \\
        \rowcolor{gray!10} - &
        GDLF-OW &
        - &
        - &
        Links from \citeauthor*{NDSS:RPJvJ18}~\cite{NDSS:RPJvJ18} &
        Random subpage &
        - &
        7e4 &
         &
         &
        7e4 &
        - &
        - \\
        \cite{SP:MHORHW23} &
        BigEnough &
        2021 &
        Web &
        Open PageRank top pages &
        Index page &
        TBB &
        3.8e4 &
        950 &
        20 &
        1.9e4 &
        On request &
         \\
        \rowcolor{gray!10} \cite{SP:DYLZLXXW23} &
        Multi-tab &
        2022 &
        Web &
        Alexa top pages &
        Index page (multi-tab) &
        TBB; Selenium &
        5.7e5 &
         &
         &
         &
        \href{https://github.com/Xinhao-Deng/Multitab-WF-Datasets}
        {Online \faExternalLink} &
        \cite{SP:DYLZLXXW23} \\
        \cite{PoPETS:JanWai23} &
        $D(\mathrm{tbs, tor})$ &
        2022 &
        Web &
        Wikipedia browsing &
        Random subpage &
        tor-browser-selenium &
        2e4 &
        98 &
        200 &
         &
        \href{https://explainwf-popets2023.github.io/data/}
        {Online \faExternalLink} &
        \\
        \rowcolor{gray!10} \cite{CCS:BaBoHo23} &
        Drift &
        Ca. 2023 &
        Web &
        Popular websites, links from
        \citeauthor*{NDSS:RPJvJ18}~\cite{NDSS:RPJvJ18} &
        Index page &
        TBB~11.0.10; tor-browser-selenium~0.6.3 &
        1.5e4 &
        90 &
        \approx 110 &
        5e3 &
        \href{https://github.com/SPIN-UMass/Realistic-Website-Fingerprinting-By-Augmenting-Network-Traces}
        {Online \faExternalLink} & \cite{CCS:BaBoHo23}
        \\
        \rowcolor{green!10} &
        \dataset{} &
        2023 &
        Any &
        Real Tor usage &
        Visited service &
        Real client software &
        1.4e7 &
        \multicolumn{3}{c}{$\langle$ \num{1.1e6} domains $\rangle$}
        &
        \href{https://doi.org/10.5281/zenodo.10620519}
        {On request \faExternalLink}
        &
        \\
        \cite{mitseva2024stop} &
        ALEXA-WSC-FG/BG &
        Ca. 2024 &
        Web &
        Alexa top sites, random subpage &
        Random subpage &
        TBB 7.5.6 &
        8.6e5 &
        9000 &
        90 &
        4.5e4 &
        No &
        \cite{mitseva2024stop}\\
        \rowcolor{gray!10} \cite{zhao2024towards} &
        CW/OW &
        Ca. 2024 &
        Web &
        Alexa top sites, random subpage &
        Random subpage (multi-tab) &
        TBB &
        8.1e4 &
        1000 &
        10 &
        9.3e3 &
        \href{https://zenodo.org/records/13252953}
        {Online \faExternalLink} &
        \cite{zhao2024towards}\\      
        \cite{CCS:MeJiJu25} &
        D1–D7 &
        2024 &
        Web &
        Tranco top sites &
        Index page &
        TBB 10.5; Chrome 112.0 &
        7.4e5 &
        100 &
        700 &
        4.00e3 &
        \href{https://zenodo.org/records/16607834}
        {Online \faExternalLink} &
        \cite{CCS:MeJiJu25}\\      
        \bottomrule
      \end{tabulary}
    \end{table}
  \end{landscape}

  \twocolumn
}
  
\end{document}